\documentclass[journal]{IEEEtran}
\usepackage{algpseudocode}
\usepackage{xcolor}
\usepackage{ulem}
\usepackage{booktabs}  
\usepackage{amsmath,amssymb,amsfonts}
\usepackage{amsthm}
\usepackage{algpseudocode}
\usepackage{algorithm}
\usepackage{mathrsfs}
\usepackage{xcolor}
\usepackage{booktabs}
\usepackage{pifont} 
\newcommand{\cmark}{\ding{51}} 
\newcommand{\xmark}{\ding{55}} 

\def\BibTeX{{\rm B\kern-.05em{\sc i\kern-.025em b}\kern-.08em
    T\kern-.1667em\lower.7ex\hbox{E}\kern-.125emX}}
\usepackage{array}
\usepackage[caption=false,font=normalsize,labelfont=sf,textfont=sf]{subfig}
\usepackage{textcomp}
\usepackage{stfloats}
\usepackage{url}
\usepackage{multirow}
\usepackage{verbatim}
\usepackage{graphicx}
\usepackage{cite}
\newtheorem{proposition}{Proposition}

\begin{document}

\title{Dynamic Control Aware Semantic Communication Enabled Image Transmission for Lunar Landing}

\author{Fangzhou Zhao,~\IEEEmembership{Graduate Student Member}, Yao Sun,~\IEEEmembership{Senior Member,~IEEE}, Jianglin Lan, and Muhammad Ali Imran,~\IEEEmembership{Fellow,~IEEE}
\thanks{*Jianglin Lan was funded by the Leverhulme Trust Early Career Fellowship (ECF-2021-517).}
\thanks{Fangzhou Zhao, Yao Sun, Jianglin Lan, and Muhammad Ali Imran are with the James Watt School of Engineering, University of Glasgow, Glasgow G12 8QQ, U.K.}
\thanks{Yao Sun is the corresponding author. (Email: Yao.Sun@glasgow.ac.uk)
}
}
\maketitle

\begin{abstract}
The primary challenge in autonomous lunar landing missions lies in the unreliable local control system, which has limited capacity to handle high-dynamic conditions, severely affecting landing precision and safety. Recent advancements in lunar satellite communication make it possible to establish a wireless link between lunar orbit satellites and the lunar lander. This enables satellites to run high-performance autonomous landing algorithms, improving landing accuracy while reducing the lander's computational and storage load. 
Nevertheless, traditional communication paradigms are not directly applicable due to significant temperature fluctuations on the lunar surface, intense solar radiation, and severe interference caused by lunar dust on hardware. The emerging technique of semantic communication (SemCom) offers significant advantages in robustness and resource efficiency, particularly under harsh channel conditions.
In this paper, we introduce a novel SemCom framework for transmitting images from the lander to satellites operating the remote landing control system. The proposed encoder-decoder dynamically adjusts the transmission strategy based on real-time feedback from the lander's control algorithm, ensuring the accurate delivery of critical image features and enhancing control reliability.
We provide a rigorous theoretical analysis of the conditions that improve the accuracy of the control algorithm and reduce end-to-end transmission time under the proposed framework. Simulation results demonstrate that our SemCom method significantly enhances autonomous landing performance compared to traditional communication methods.

\end{abstract}


%
\IEEEpeerreviewmaketitle

\section{Introduction}
Over the past decade, an increasing number of countries have embarked on a new wave of lunar exploration initiatives, with autonomous lunar landing missions playing a crucial role in enabling these efforts\cite{cnlanding,indialanding, UAE, lin2024return,r3-6,r3-7}.
However, some failures of autonomous lunar landing missions indicate that deploying a reliable local autonomous landing control system in a highly dynamic environment presents significant challenges\cite{india3,fail1}.
To improve the reliability of autonomous lunar landing systems, recent studies have explored the use of advanced machine learning algorithms, particularly reinforcement learning (RL), for image-based decision-making and closed-loop descent control\cite{imagelunar1}. RL has demonstrated promising capabilities in coping with high-dimensional visual observations and partially observable dynamics, making it more suitable than classical controllers such as Proportional–Integral–Derivative (PID) and Model Predictive Control (MPC) for real-time landing under uncertain lunar environments\cite{imagelunar2}.
However, due to the strict size and weight limitations on lunar landers, incorporating sufficient computing resources to support real-time RL inference presents considerable design challenges.\cite{cost}.

Recent advancements in lunar communication satellites have made it possible for landers to establish a wireless communication link with lunar orbit satellites\cite{relay1,navi1,r4-1,r4-2,r4-3,r4-4,r4-5,76,77,78}. This prompts a different angle to investigate the lunar landing mission, hypothesizing a lunar lander is remotely controlled via a high-performance lunar orbiting satellite.
Specifically, lunar surface images collected by the lander are transmitted to a lunar orbiting satellite that possesses sufficient computational resources for image processing.
Based on the images received, the satellite then runs an automatic control algorithm to generate appropriate control decisions for the lander maneuver. 
These control decisions are sent back to the lander to execute. The control actions determined by satellites with sufficient computing resources can help to achieve more precise control performance than those determined locally on the lander itself. 

However, considerable challenges tied to the hypothesis arise from the image transmissions between the lander and satellites.
On the lunar surface, the temperature variation between day and night can exceed 300°C, which may cause huge fluctuations in the noise incurred by the lander communication hardware, leading to very unstable wireless links\cite{lunarsnr,raza2022toward}.
Furthermore, the intense solar radiation and lunar dust present on the lunar surface can significantly interfere with the wireless channel, leading to severe signal distortion and transmission latency.
In addition to these impairments, lunar missions inherently face long propagation delays between the satellites and the lunar surface and the high bit error rate in deep-space links. These deep-space specific challenges further highlight the need for advanced communication paradigms tailored to the lunar landing task\cite{toyoshima2021recent,10988568,al2022survey,r3-1,r3-2,r3-3,r3-4,r3-5}.
Therefore, traditional communications may cause the satellite to receive a distorting key feature in the lunar surface image and produce inaccurate control actions. 
In image-based autonomous lunar landing, the significance of each pixel varies across different regions of the image\cite{imagelunar2}. For instance, distortions in critical features near the landing site, such as craters, can lead to incorrect navigation data, causing the lander to move in the wrong direction. 
This is especially problematic for RL based real-time control algorithms for lunar landing, which highly rely on the accuracy of key features extracted from the input image and wrong navigation information will directly lead to wrong action output\cite{landing,imagelunar2}.
Therefore, designing a wireless data transmission scheme that accurately transmits the key feature in the image to support the lunar landing control is strongly needed.

The emerging technique of semantic communication (SemCom) has shown promise in accurately transmitting key image features. 
Instead of transmitting every bit, SemCom focuses on preserving meaning by leveraging machine learning to extract and convey the essential semantics of the source information\cite{zhao2024enhancing,chengsi,liu2024knowledgeassistedprivacypreservingsemantic,10742565,r2-1,r2-2,r2-3,r2-4,r2-5}. 
These semantics preserve task-relevant features while discarding redundant or low-impact information. Although semantic information is condensed, the decoder leverages contextual redundancy and the prior knowledge embedded in deep learning models to reconstruct meaningful content even under bit-level distortions\cite{deepsc,toVQA, sc7,sc8,r2-6,r2-7,r2-8,r2-9,r2-10,79,80}. This enables SemCom to tolerate transmission errors, making it particularly suitable for noisy or interference-prone lunar environments.
Therefore, using SemCom to ensure the transmission of the key semantics (the key input features required by the control system) for the lunar landing control system can enhance the satellite-controlled autonomous lunar landing capabilities.
However, one of the key challenges is how to design a SemCom transceiver that is aware of the semantics required by the control task and the potential stability constraints. 
Although high-performance SemCom transceivers are capable of accurately extracting key semantic information, their encoding and decoding process extends the end-to-end transmission time, inducing additional control delays and compromising control stability. 
To reduce the encoding and decoding time, the implementation of lightweight SemCom transceivers is advisable. However, such transceivers may not ensure precise semantic extraction.
Therefore, it is necessary to design a SemCom transceiver that can dynamically adjust the transmission strategy according to the real-time feedback of the lander control algorithm, thereby reducing encoding and decoding time while ensuring control stability.

To address the above challenges, we propose a dynamic control-aware SemCom framework for autonomous lunar landing. 
In this framework, we design an end-to-end SemCom transceiver (named DynaSC) performing image transmissions between the lunar orbiting satellite and the lunar lander. 
The transceiver dynamically adjusts the SemCom coding strategy such that the lander transmits accurate key features of the image at the critical moments.
The main contributions of this paper are summarized as follows:
\begin{itemize}
\item
\textit{Landing control-aware SemCom framework:} 
We consider a resource-constrained image-based wireless lunar lander control scenario and establish a landing control-aware SemCom framework. The lunar lander senses image information and transmits it to the satellite through wireless SemCom, and the satellite selects actions through RL to remotely manipulate the lunar lander.
\item
\textit{Landing control-aware SemCom transceiver:} 
We design a SemCom transceiver that leverages a neural network to encode and decode key regions in images, ensuring accurate transmission while simplifying the encoding of non-essential regions to reduce overall encoding computation time. Furthermore, the SemCom transceiver dynamically adjusts the key region boundaries based on real-time reward feedback from the lander's RL control algorithm, thereby reducing image encoding and decoding time while maintaining control stability.
\item
\textit{Theoretical analysis of performance:} 
We mathematically derive the channel condition requirements for SemCom under a lunar communication model that includes signal reflections and multipath propagation caused by surface features such as craters and ridges, to achieve improved end-to-end transmission time and energy efficiency compared to traditional communication methods. Additionally, we discuss the channel conditions under which SemCom enhances the average reward expectations in RL-based lander control algorithms.
\item
\textit{Numerical simulation validation:} 
We perform numerical simulations of the proposed satellite-controlled lunar landing framework, where sensed images are transmitted using the proposed SemCom transceiver, along with two benchmark methods for comparison. The simulations are conducted under both AWGN and Rician fading channels to evaluate performance under realistic lunar conditions.
The results demonstrate that the proposed SemCom transceiver enhances the lunar lander's autonomous landing capabilities while reducing end-to-end transmission time.
\end{itemize}

The rest of the paper is organized as follows. Section II provides a review of the related works. Section III presents the control-aware SemCom framework, followed by the control-aware SemCom transceivers in Section IV. The properties analysis of the landing control-aware SemCom is conducted in Section V. Section VI reports the numerical simulations and Section VII draws the conclusions. 

\section{Related Works}
\begin{table*}[t]
    \setlength{\abovecaptionskip}{-3pt}
    \setlength{\belowcaptionskip}{-10pt}
    \vspace{-1mm}
    \small
    \caption{Comparison of the proposed method with existing transceiver designs for image SemCom}
    \begin{center}
    \renewcommand{\arraystretch}{1.15}
    \setlength{\tabcolsep}{5pt}
    \begin{tabular}{c m{4.5cm} >{\centering\arraybackslash}m{1.15cm} >{\centering\arraybackslash}m{1.4cm} >{\centering\arraybackslash}m{1.2cm} >{\centering\arraybackslash}m{1.2cm} >{\centering\arraybackslash}m{1.4cm} >{\centering\arraybackslash}m{2cm}}
        \toprule[1.2pt]
        \textbf{Reference} & \centering \textbf{Design Purpose} & \textbf{Compute Offload} & \textbf{Lightweight Model} & \textbf{Task-oriented design} & \textbf{Control-aware design} & \textbf{Dynamic Coding} & \textbf{Performance Metric} \\
        \midrule
        \cite{sc1}   & Robustness to semantic noise via adversarial training and feature masking & \xmark & \cmark & \xmark & \xmark & \xmark & Classification accuracy \\
        \cite{sc2}   & Semantic compression for image data size reduction & \xmark & \xmark & \xmark & \xmark & \cmark &FID, KID and mIoU \\
        \cite{sc3}   & Semantic preservation via shared knowledge libraries & \xmark & \xmark & \xmark & \xmark & \xmark & PSNR \\
        \cite{sc4}   & Task success maximization via adaptive compression and resource allocation & \xmark & \xmark & \cmark & \xmark & \cmark & PSNR, SFI and KID \\
        \cite{sc5}   & Multi-task AI support via semantic aware deep JSCC & \cmark & \xmark & \cmark & \xmark & \xmark & Classification accuracy \\
        \cite{runze} & AIGC efficiency improvement with workload-adaptive SemCom & \cmark & \xmark & \cmark & \xmark & \xmark & Satisfaction score \\
        \cite{cococo} & Joint control-communication cost reduction in MLD system & \xmark & \xmark & \cmark & \cmark & \xmark & Control Cost \\
        \cite{10431795} & Unified semantic system for multi-task efficiency. & \xmark & \cmark & \xmark & \xmark & \cmark & Classification accuracy \\
        \cite{10945983} & Bandwidth efficient semantic transmission with diffusion models. & \xmark & \xmark & \xmark & \xmark & \xmark & PSNR, SSIM, CLIP and FID \\
        \cite{10599525} & Low-latency semantics transmission using foundation models. & \xmark & \xmark & \cmark & \xmark & \xmark & CLIP, BER \\
        \cite{r2-4} & Adaptive SemCom network resource allocation. & \cmark & \xmark & \xmark & \xmark & \cmark &  SC-QoS,SQE\\
        \cite{r2-8} & Knowledge-Graph SemCom robust coding. & \xmark & \xmark & \cmark & \xmark & \xmark & Token Similarity \\
        \textbf{Our work} & Precision lunar landing via SemCom with control feedback & \cmark & \cmark & \cmark & \cmark & \cmark & Task Success Rate \\
        \bottomrule[1.2pt]
    \end{tabular}
    \end{center}
\end{table*}
To support accurate autonomous lunar landings, numerous researchers have made efforts on developing image-based lunar surface navigation technologies or autonomous landing control systems.
Previous studies have demonstrated the effectiveness of image-based autonomous landing techniques on the lunar surface\cite{imagelunar1,imagelunar2,imagelunar3,silvestrini2022optical}. 
An image-based reinforcement meta-learning is proposed in \cite{imagelunar1} for solving the lunar pinpoint powered descent and landing task with uncertain dynamic parameters and actuator failure. This method can achieve low error guidance and landing when the environment is partially observed and the state of the spacecraft is incompletely known.
The research \cite{imagelunar2} studies a navigation method based on a convolutional neural network to achieve a precise lunar landing mission. This method calculates the relative position and velocity of the lander by matching the features of the observed craters. The study \cite{imagelunar3} introduces an autonomous landing method based on visual odometry without an onboard map of the lunar surface. The advantage of this technique is that the lander can autonomously navigate and land using only images observed by an onboard monocular camera, without the need for an onboard map. This research \cite{silvestrini2022optical} proposes a vision navigation system for lunar pinpoint landing, where crater detection and feature matching provide pseudo-measurements that are fused with altimeter data through an Extended Kalman Filter to achieve absolute navigation at the Moon’s South Pole.

Based on the aforementioned studies, image-based autonomous lunar landing has been widely explored and is expected to play a critical role in future missions. However, these methods often rely on large-scale neural networks, posing significant computational burdens that are difficult to accommodate on resource-constrained landers. A promising alternative is to offload the computation by transmitting observation images to a more capable lunar-orbiting satellite for processing and remote control.
While several proposals have expected building dedicated lunar communication infrastructures\cite{gate,lunarnet,parker2022lunar,73,74,75}, these systems remain in the planning stage and are not yet available for practical deployment. As a result, current missions still rely on traditional satellite relays, which offer limited bandwidth and are vulnerable to interruption. These constraints make it difficult to ensure continuous, high-reliability communication for vision-based landing control, especially when large image data needs to be transmitted in real time.

Fortunately, SemCom technologies provide the possibility to improve the communication reliability of satellites and landers. 
Several studies have focused on designing SemCom transceivers specifically for image transmission, demonstrating notable improvements in robustness and transmission efficiency\cite{sc1,sc2,sc4,10599525,li2025goalorientedsemanticcommunicationwireless,11085101,r2-4,r2-8}.
The study \cite{sc1} introduces a robustness SemCom against semantic noise, which was achieved by incorporating semantic noise samples into the training dataset for adversarial training, thereby improving the noise resilience of the SemCom transceiver.
The research \cite{sc2} proposes a semantic coding method to guarantee the key semantics at a poor bit rate, which used a generative adversarial network to reconstruct the key features. 
The research \cite{sc4} proposes an end-to-end SemCom system for image transmission that leverages shared semantic knowledge libraries, using a classifier and dictionary learning for feature extraction at the sender, and a diffusion model-based generator for reconstruction at the receiver, with a semantic fidelity index (SFI) to evaluate reconstruction quality. The study \cite{10599525} develops a latency-aware SemCom framework using generative AI models, employing multi-modal semantic decomposition and adaptive transmission schemes to achieve ultra-low-rate, high-fidelity signal synthesis under dynamic channel conditions. The study \cite{li2025goalorientedsemanticcommunicationwireless} proposes a stable diffusion–based SemCom framework that extracts keyframe semantics and denoises channel noise, achieving superior video quality and robustness under both known and unknown channels. 
This paper \cite{r2-4} proposes an adaptive semantic resource allocation paradigm with semantic-bit quantisation and optimizing SC-QoS via hybrid DRL for superior robustness and efficiency.
The study \cite{r2-8} proposes a knowledge graph–based SemCom framework that aligns entities with messages and enhances robustness under noisy channels.
The study \cite{11085101} proposes an edge Cognitive SemCom Agent that uses large AI models to align modalities and generate personalized multimodal communication policies, significantly improving accuracy, intent satisfaction, and latency.

Beyond general image reconstruction, an increasing number of studies have explored task-oriented SemCom designs, where semantic encoding is explicitly optimized to support specific downstream tasks under resource constraints\cite{sc3,runze,sc5,10431795,10945983}.
The study \cite{sc3} proposes a deep joint source and channel coding (JSCC) framework that simultaneously supports multi-task services, including image data recovery and classification, by maximizing coding rate reduction and minimizing mean square error (MSE), while incorporating a gated design to improve robustness against variational wireless channels.
The research \cite{sc5} proposes a task-oriented communication architecture with an adaptable semantic compression (ASC) method to optimize semantic transmission, introducing CRRA and CRRAUS algorithms for efficient compression ratio and resource allocation, achieving significant data size reduction and improved task success probability in delay-intolerant scenarios.
The research \cite{runze} proposes a SemCom framework for transmitting images in AI-generated content (AIGC) tasks, which creatively integrated the diffusion model in the semantic encoder and decoder, thereby transmitting the intermediate features of AI-generated content.
The study \cite{10431795} proposes U-DeepSC, a unified SemCom system that dynamically adjusts transmitted features across tasks and channel conditions via a lightweight feature selection module and a shared codebook, achieving task-specific performance while reducing transmission overhead and model size.
The research \cite{10945983} proposes a diffusion-based SemCom system with variational autoencoder compression, significantly improving reconstruction quality while reducing bandwidth requirements compared to traditional methods.

Although recent studies have adopted advanced architectures such as DeiT-based and diffusion-based models to improve SemCom compression efficiency under resource constraints, they typically focus on optimizing semantic representation or reconstruction accuracy under a fixed task. 
However, they do not incorporate task-specific control feedback into the transmission process. 
In particular, their sparsification strategies are either static or determined by heuristic criteria, which limits their adaptability in dynamic control scenarios such as autonomous lunar landing.

An initial attempt to combine semantics and control is found in \cite{cococo}, which focuses on symbolic logic-level communication within mixed logical dynamical systems. However, it does not support real-valued features such as images or address the challenges of vision-based control under constrained conditions. In contrast, our work proposes a control-aware sparsification mechanism that dynamically adjusts semantic transmission based on the lander’s reward signal. This enables adaptive prioritization of control-relevant features and tightens the integration between communication and control—essential for mission-critical tasks with strict bandwidth and latency requirements.

\section{System Model}
\begin{table}[ht]
\centering
\caption{List of Symbols}
\begin{tabular}{ll}
\hline
\textbf{Symbol} & \textbf{Meaning} \\
\hline
$S, A, Z$ & State space, action space, and observation space \\
$\Phi, \Omega$ & State transition and observation probability functions \\
$R$ & Reward function \\
$X, Y$ & Observable lander states and latent lunar surface states \\
$w_t$ & Sensed image at time $t$ \\
$\hat{w}_t$ & Reconstructed image after transmission and decoding \\
$\eta(\cdot)$ & Neural network to extract semantic feature\\
$y_t$ & True semantic state of the lunar surface at time $t$ \\
$z_t$ & Noisy semantic observation at time $t$ from $\hat{w}_t$ \\
$x_t$ & Observable physical state of the lander at time $t$ \\
$u_t$ & Control action at time $t$ \\
$b_t$ & Belief state at time $t$ \\
$\pi$ & Policy mapping belief to action \\
$V^\pi(b_t)$ & Value function under policy $\pi$ at belief $b_t$ \\
$\delta_t$ & Proportion of semantically important image patches \\
$\kappa$ & Normalization coefficient for $\delta_t$ \\
$E_e(\cdot), E_d(\cdot)$ & Semantic encoder and decoder networks \\
$s_t, \hat{s}_t$ & Semantic symbol and its after transmission version \\
$h(t)$ & Rician fading channel coefficient \\
$n$ & Additive White Gaussian Noise \\
$\gamma$ & Instantaneous SINR \\
$p_\gamma(\gamma)$ & PDF of $\gamma$ under Rician fading \\
$I_0(\cdot)$ & Zeroth-order modified Bessel function of the first kind \\
$T_c(w_t)$ & Computation time for processing $w_t$ \\
$T_d$ & Transmission time \\
$Z(w_t)$ & Bit size of source image \\
$Z(s_t)$ & Bit size of semantic representation \\
$C$ & Channel capacity \\
$E_c(w_t), E_d$ & Computation and transmission energy consumption \\
$L_{\text{MSE}}$ & Mean squared error loss \\
$L_{\delta_t}$ & Loss term for matching sparsity ratio $\delta_t$ \\
$L_{\text{distill}}$ & Knowledge distillation loss \\
$L$ & Overall training loss \\
\hline
\end{tabular}
\end{table}

\begin{figure*}[htp]
	\centering
	\includegraphics[width=0.9\textwidth]{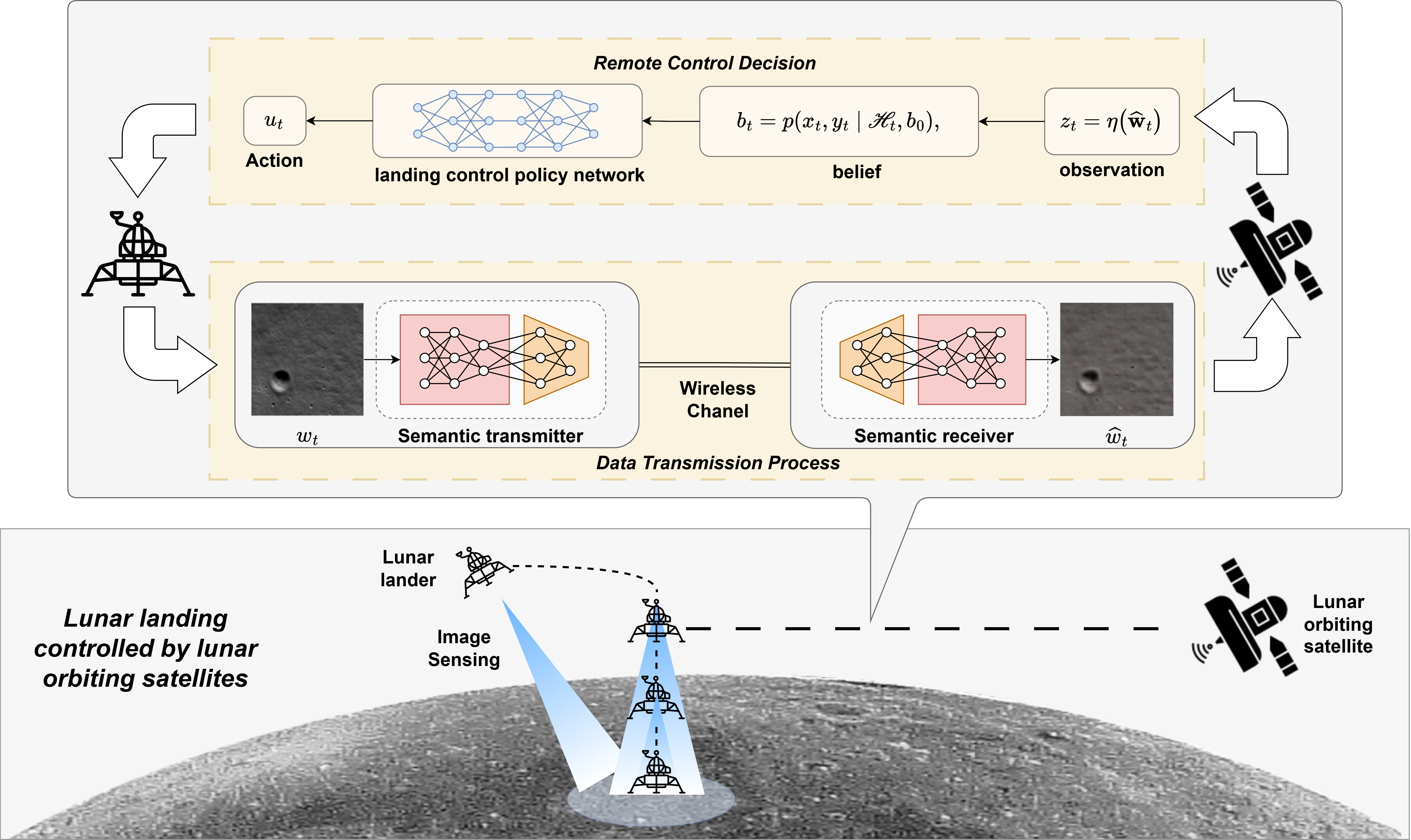}
	\caption{The proposed RL-based lunar landing control framework.}
	\label{img1}
\end{figure*}
We consider a lunar landing control system consisting of a satellite and a wireless remote-controlled lander. As shown in Fig. 1, the lander first observes the lunar surface and transmits the observed images to the satellite via wireless communication. After receiving the observation data, the satellite makes control decisions for the lander's action and sends them to the lander through wireless communication.  
In this framework, the full state of the lunar landing environment—including precise lander dynamics and surface terrain—is only partially observable, and must be inferred from the images transmitted over the noisy communication channel. As a result, it is necessary to incorporate both observation uncertainty and the effects of SemCom into the control model.

\subsection{RL-based Landing Control Model}
For this incompletely observed environment, we formulate it as a Partially Observable Markov Decision  Process (POMDP) $(\mathcal S,\mathcal A,\mathcal Z,\Phi,\Omega,R)$, where $\mathcal S,\mathcal A,\mathcal Z$ represent states, actions, and observations, respectively, $\Phi$ and $\Omega$ are the probability distribution of state transitions and observations, and $R$ is the reward function.

In our model, the state space $\mathcal S=(\mathcal X,\mathcal Y)$ is a finite set, where $\mathcal{X}$ denotes the fully observable states of the lander, and $\mathcal Y$ corresponds to the partially observable lunar surface state. We consider a total time $T$ and $x_t \in \mathcal{X}$ and $y_t \in \mathcal Y$ represent the states at time $t \in T$.
In particular, we use a neural network $\eta(\cdot)$ to extract $y_t$ from the sensed image ${\mathbf w}_t$ (the source information), i.e., $y_t=\eta (\mathbf w_t)$.

Similar to \cite{imagelunar1}, we consider a general lunar landing scenario,  where the action can be the lander's propulsion direction and thrust magnitude. In this work, we do not specify the action but generally denote the action at time $t$ as $u_t \in \mathcal A$.

The observation space $\mathcal Z$ corresponds to the observed state of the lunar surface based on the received images.
Since the sensed images $\mathbf w_t$ is transmitted to the satellite by SemCom over noisy wireless channels, the received lunar surface image $\mathbf {\hat w}_t$ is fed to the neural network for generating navigation data to obtain the noisy observation of the lunar surface state, i.e, $z_t=\eta (\mathbf {\hat w}_t)$, where $z_t \in \mathcal Z$.

For a general fully observed MDP problem, the state transition probability of the fully observed state $x_t$ can be modelled in a deterministic manner $x_{t+1} = f(x_{t_1},u_{t_1})$, where $f(\cdot)$ is the transition function to determine the next state given the current state and action. The state transition probability of $y_t$ can be modelled as the distribution function $\Phi(y_t)=p(y_t|u_{t-1},y_{t-1})$.
For the considered POMDP problem, since the true state $y_t$ cannot be directly observed, the belief $b_t$ is introduced to estimate the confidence of $y_t$ under the observation $z_t$. The belief $b_t$ is the probability of the current state $(x_t,y_t)$ given historical observations, i.e., 
\begin{equation} 
b_t = p(x_t, y_t|\mathcal H_t,b_0),
\end{equation}
where $\mathcal H_t$ is the observation and action history denoted as ${\mathcal H_t=\{z_1,a_1,z_2,a_2...,z_{t-1},a_{t-1}\}}$ and the $b_0$ is an initial belief. The update for the belief follows
\begin{equation} 
b_{t} = \tau(b_{t-1}, u_{t-1}, z_{t}),
\end{equation}
where $\tau(\cdot)$ is the belief update function.
Based on the belief $b_t$, the landing control policy $\pi$ decides the action $u_t$ of the lander, denoted as
\begin{equation} 
\label{u}
u_t=\pi(b_t).
\end{equation}

To find an optimal policy, we define a reward function $R=r(x_{t-1},y_{t-1},u_t,x_t,y_t)$ for taking action $u_t$ under belief $b_t$. The optimal policy $\pi^*$ is denoted as
\begin{equation} 
\pi^*=\underset{\pi\in\Pi}{\arg\max}\mathbb{E}\bigg[\sum\limits_{t=0}^T r(x_{t-1},y_{t-1},u_t,x_t,y_t)|b_0\bigg].
\end{equation}
Given $b_t$, we can get the probability distribution of observation $\Omega(z_t)=p(z_t|x_{t},y_{t},u_{t-1})$. The estimation of the observation can be given by a distribution that follows
\begin{equation}
\begin{aligned}
p(z_t|b_t,u_t)=\sum \Omega(z_t)\sum {\Phi(y_t) b_{t-1}}.
\end{aligned}
\end{equation}
Based on the estimation of the observation, we can write the belief-based state value function corresponding to $\pi^*$ as
\begin{equation}
\label{v}
V^{\pi^*}(b_t) \!=\! \max_{u_t\in \mathcal{A}} \left[r(b_t,u_t) \!+\! \gamma\sum_{z_t\in \mathcal Z}p(z_t|b_t,u_t)V^*(b_{t-1}) \right]. \!\!
\end{equation}
The objective is to achieve $V^{\pi^*}(b_t)$ by finding the optimal control policy $\pi^*$.

\subsection{SemCom Model}
We exploit SemCom to transmit sensed lunar surface image $\mathbf {w}_t$. The terrain category, spatial arrangement, and visual characteristics are implicit semantics within the image, which in our framework refer to high-level features such as task-critical regions (e.g., craters or landing zones) and latent embeddings extracted by deep neural networks that capture these structures for reliable control\cite{sc2}. Typically, in high-performance neural networks for image processing (such as Vision Transformers (ViT)\cite{ViT}), the image is divided into several small patches for encoding and decoding. Each image patch has a different importance to the overall image on semantics\cite{sc1}. Thus, the image $\mathbf {w}_t$ can be represented as $\mathbf {w}_t=\{w_{t1},w_{t2},...,w_{tN}\}$, where $w_{tk}$ denotes the $k$-th patch of the $\mathbf {w}_t$. 

To ensure the accurate transmission of $\mathbf {w}_t$ via SemCom, we optimize the encoding process based on the semantic importance of each patch.
In particular, the patches with high semantic significance in $\mathbf{w}_t$ are prioritized for accurate transmission, while other patches can undergo simplified encoding to accommodate the limited bandwidth and further reduce computing time.
Meanwhile, the proportion of patches with high semantic importance, denoted as $\delta$, varies with the lander state $x_t$ and the environment state $y_t$.
Intuitively, a small $\delta$ can reduce the number of bits being transmitted, thereby reducing the energy and the information transmission delay. 
However, overly aggressive compression may degrade image accuracy, increasing the mismatch between the observation $z_t$ and the true environment state $y_t$, which can negatively affect landing decision quality. Therefore, $\delta$ directly influences the trade-off between communication efficiency and control performance.

To maintain the landing control accuracy while reducing computation time, we should dynamically adjust $\delta$ according to the real-time derivative of the reward within RL algorithms. Proposition 1 reveals the relation between $\delta$ and the real-time derivative of the reward. 
\begin{proposition}
Assuming that the observation error caused by reduced $\delta$ follows a normal distribution independent of time.
Let $z_{t1}$ and $z_{t2}$ be the observations at times $t_1$ and $t_2$ respectively, and $\sigma_{t_1}$ and $\sigma_{t_2}$ be the variances of the observed distributions at these times.
Given that
$$\frac{{{\partial }}{r(b_{t_1},u_{t_1})}}{\partial u_{t_1}}>\frac{{{\partial }}{r(b_{t_2},u_{t_2})}}{\partial u_{t_2}},$$ the rate of change of variance satisfies  
$$\frac{{\epsilon}^2}{\sigma_{t_1}+{\epsilon}^2}<\frac{{\epsilon}^2}{\sigma_{t_2}+{\epsilon}^2}.$$
\end{proposition}
\renewcommand\qedsymbol{$\blacksquare$}
\begin{proof}
See Appendix A.
\end{proof}

At time $t$, the proportion of patches with high semantic importance $\delta_t$ is defined as
\begin{equation}
\begin{aligned}
{{\delta }_{t}}=\frac{1}{1+e^{\kappa\frac{{{\partial }}{r(b_t,u_t)}}{\partial u_t}}},
\end{aligned}
\end{equation}
where $\kappa$ is the normalized coefficient. 

For a SemCom-based image transmission, $\mathbf {w}_t$ is first encoded to bits representing semantics ${\mathbf {s}}_{t}$ by a landing control-aware semantic encoder, as follows:
\begin{equation} 
{\mathbf {s}}_{t} = {\mathcal{ E}_e}\left ({{\mathbf {{w_{t}}}}; { \boldsymbol \alpha, \delta_t}} { }\right),
\end{equation}
where ${\mathcal{ E}_e}$ is the neural network of the encoder with the trainable parameters $\boldsymbol \alpha$.

The transmission takes place over a lunar wireless channel with complex terrain-induced characteristics. Crater walls and surface irregularities introduce strong reflections and multipath propagation, which lead to signal dispersion and fading \cite{santra2020radio}. To model these effects, we adopt a Rician fading channel representation \cite{divsalar2022comparing}.

After ${\mathbf {s}}_{t}$ are transmitted through the noisy physical channel, the received symbols ${\mathbf {\hat s}}_{t}$ are
\begin{equation} 
{\mathbf {\hat s}}_{t} = {\mathbf h(t) *{\mathbf {s}}_{t}} + {\mathbf {n}},
\end{equation}
where $\mathbf h(t)$ is the complex channel coefficient vector modeled as a Rician fading process, capturing both the deterministic LoS and stochastic multipath components of the lunar terrain-induced channel. The term $\mathbf {n}$ is the additive white Gaussian noise (AWGN) of the channel.
Correspondingly, the reconstructed image is finally represented as
\begin{equation} 
{\mathbf {\hat w}}_{t} = {\mathcal{ E}_d}\left ({{\mathbf {\hat s}}_{t}; { \boldsymbol \alpha, \delta_t}} {}\right),
\end{equation}
where ${\mathcal{ E}_d}$ is the neural network of the landing control-aware semantic decoder.
\begin{figure*}[htp]
	\centering
	\includegraphics[width=0.8\textwidth]{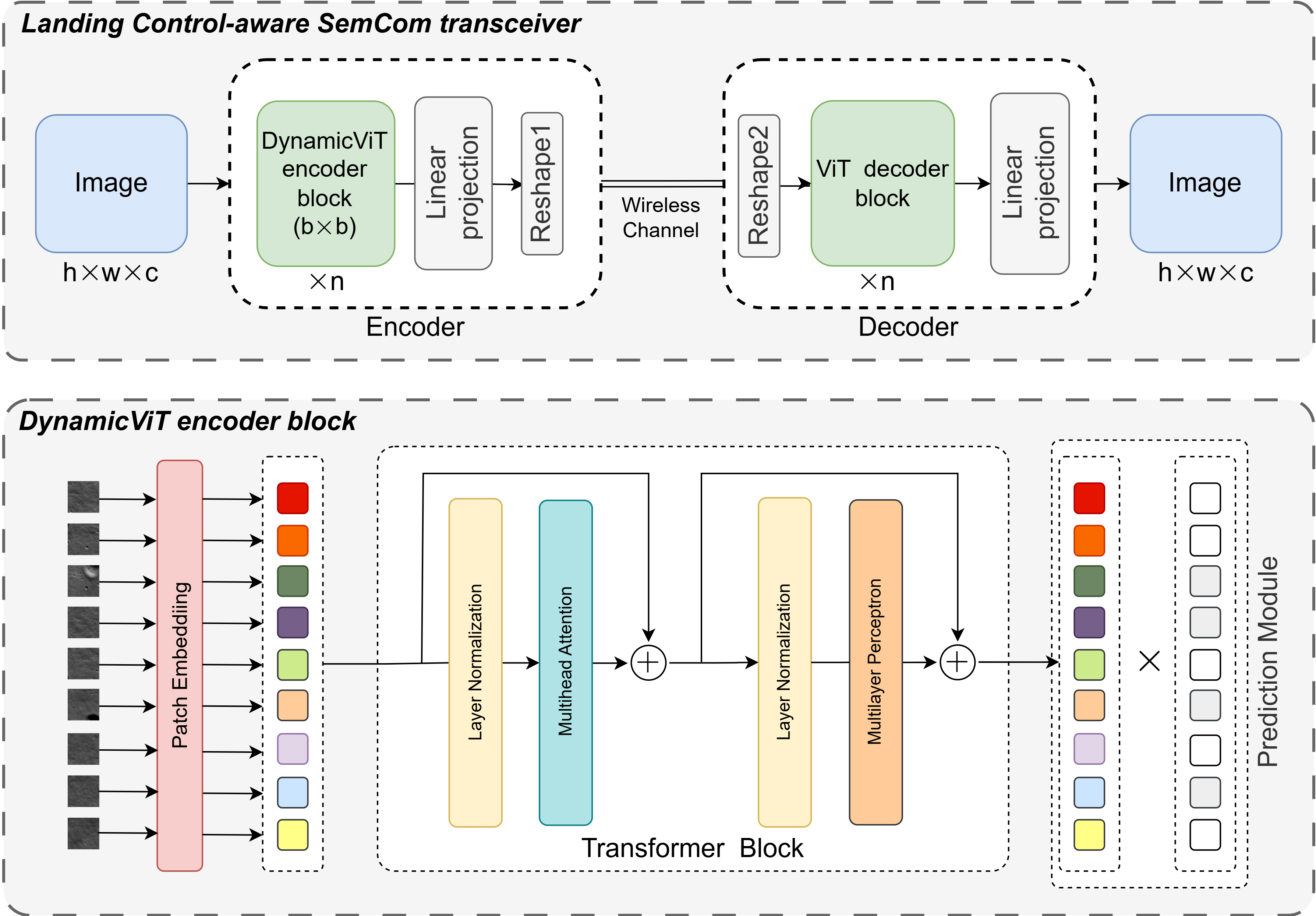}
  	\caption{Dynamic sparsification ViT-based SemCom encoder and decoder network.}
	\label{img2}
\end{figure*}

\section{Landing Control-aware SemCom Transceiver}
In this section, we design the SemCom transceiver (named DynaSC) to dynamically adjust the coding strategy according to the derivative of the lander control algorithm reward.
DynaSC can recover the semantic information in images from neighbouring patches through a training framework similar to Masked AutoEncoder (MAE) \cite{MAE}. Meanwhile, the importance of each patch is evaluated by the DynamicViT \cite{dn}, thereby the transmission strategy can be adjusted according to $\delta$.

\subsection{Semantic Encoder}
As shown in Fig. 2, the proposed semantic encoder consists of $n$ DynamicViT modules. Each DynamicViT module preserves all the structure of the traditional ViT but with an additional prediction module at its end. In particular, the input image $\mathbf x_t \in \mathbb R^{h \times w \times c}$ is first divided into $N$ patches of size $b \times b$. Then each patch $w_{tk}$, $k \in [1,N]$, is pulled as a one-dimensional vector and its dimension is compressed by linear projection to obtain the token $\mathbf t_{tk} \in \mathbb R^{{L_s} \times 1}$, where $L_s$ is the token length. 

All tokens are processed sequentially through a transformer block, enabling global attention computation and feature extraction. Specifically, the output features from the transformer block are passed to a prediction module, which is a lightweight multilayer perceptron network that assesses the importance of each token for the original image. 
The prediction module adjusts the number of high-importance tokens to be transferred to the next ViT block based on a dynamic \(\delta_t\), which is computed in real time according to the gradient of the lander control algorithm reward. This mechanism enables the transmission strategy to focus on control-relevant tokens, other tokens will compute attention only with themselves, rather than with all tokens, thereby reducing the overall computational cost.

\subsection{Semantic Decoder}
The decoder of the proposed DynaSC consists of $n$ ViT decoder blocks, each contains a transformer module. The ViT decoder block receives all tokens to reconstruct the input image by predicting the pixel value represented by each token. Noisy regions in the image can be partially restored by the contextual information from neighbouring areas, thus reducing the noise impact on overall image semantics. The last decoder layer is a linear projection layer that reshapes the vector into $b \times b$ patches and reconstructs the input image $\mathbf {\hat w}$. In DynaSC, we use the average mean squared error between the original input image $\mathbf w_{t}$ and the reconstructed image $\mathbf {\hat w}_{t}$ as the loss function 
\begin{equation}
\begin{aligned}
\mathcal {L}_{MSE} = \frac {1}{N}\sum _{i=1}^{N} \|{\mathbf w_{t}-  { \mathbf {\hat w}_{t}}}\|^2.
\end{aligned}
\end{equation}

\subsection{Training Process}
To enhance the accuracy of reconstructed images in SemCom, we apply a masking technique during model training \cite{MAE}. The key idea is to input only a subset of patches into the encoder while masking the remaining patches. The decoder then reconstructs the entire image using information from the visible patches.
This approach utilizes masked patches to simulate information loss due to noise, thereby training the model to be robust against noise and enhancing its capacity to reconstruct complete images.

In the overall training of the model, the prediction module is trained in conjunction with the entire transceiver.
To enhance the stability of sparse-token training, we incorporate a knowledge distillation strategy consistent with the DeiT framework \cite{DeiT}, where a pre-trained dense semantic encoder serves as the teacher model. This distillation technique is widely adopted in sparse model training to improve convergence and maintain semantic consistency during token pruning. Knowledge distillation uses a semantic encoding network that is fully trained without sparsification, the teacher model to assist DynaSC training, with the same loss function as follows: 
\begin{align}\label{loss}
 \mathcal L ={}& {{\mathcal L}_{MSE}} + {{\lambda }_\mathbf {KL}}\mathbf {KL}(\mathbf {\hat  w_t}||\mathbf {\hat w_t}') + {\lambda }_\mathbf {\delta_t}{{\mathcal L}_{{\delta_t}}} \nonumber\\
&+ {{\lambda }_{distill}}\frac{\sum_{t=1}^{T}{\sum_{k=1}^{N}{\mathbf {\hat D}_{k}^{t,M}}}{{({{\mathbf t}_{tk}}-\mathbf t'_{tk})}^{2}}}{\sum_{t=1}^{T}{\sum_{k=1}^{N}{\hat{\mathbf D}_{k}^{t,M}}}},
\end{align}
where $\mathbf {KL}(\mathbf {\hat w_t}||\mathbf {\hat w_t}')$ is the KL divergence of $\mathbf w_t$ and the teacher model output $\mathbf {\hat w_t}'$. ${\mathbf {\hat D}_{k}^{t,M}}$ is a binary variable, indicating whether the input sample at time $t$ retains the $k$ token at the $M$ sparsification stages. ${\lambda }_\mathbf {KL}$, ${\lambda }_{distill}$ and ${\lambda }_\mathbf {\delta_t}$ are the weighting coefficients. ${\mathcal L}_{\delta_t}$ is defined as
\begin{equation}
\begin{aligned}
{{\mathcal L}_{{\delta_t}}}=\frac{1}{TM}\sum\nolimits_{t=1}^{T}{\sum\nolimits_{m=1}^{M}}\left({\delta_t}-\frac{1}{N}\sum\nolimits_{k=1}^{N}{\mathbf {\hat D}_{k}^{t,m}} \right).
\end{aligned}
\end{equation}
The training process of DynaSC is shown in Algorithm \ref{alg1}. 
\begin{algorithm}[t]
    \caption{Training Process}
    \label{alg1}
    \begin{algorithmic}[1]
        \State Initialization: Training data $\mathcal W$, learning rate $\gamma$.     
        \Repeat
        \State \parbox[t]{\dimexpr\linewidth-\algorithmicindent}{Input: Training data $\mathbf{w}_t$, actual lunar surface state $z_t$, lunar lander state $x_t$, lunar lander action $u_t$, and learning rate $\lambda$.}
        \State \parbox[t]{\dimexpr\linewidth-\algorithmicindent}{Calculate the proportion of the retained token $\delta$ based on (3) and (8).}
        \State \parbox[t]{\dimexpr\linewidth-\algorithmicindent}{Embedding the $\mathbf w_t \to \{\mathbf s_{t1},\mathbf s_{t2},...,\mathbf s_{tn}\}$.}
        \State \parbox[t]{\dimexpr\linewidth-\algorithmicindent}{$\mathbf w_t \to {\mathcal{ E}_e}\left ({{\mathbf {{w_{t}}}}; { \boldsymbol \alpha, \delta_t}} { }\right) \to \mathbf s_t \to \mathbf {\hat s}_t \to {\mathcal{ E}_d}\left ({{\mathbf {\hat s}}_{t}; { \boldsymbol \alpha, \delta_t}} {}\right) \to \mathbf {\hat w}_t$.}
        \State \parbox[t]{\dimexpr\linewidth-\algorithmicindent}{ Calculating the MSE loss function $\mathcal {L}_{MSE} = \frac {1}{N}\sum _{i=1}^{N} d\left ({\mathbf w_{t}, { \mathbf {\hat w}_{t}}}\right).$}
        \State \parbox[t]{\dimexpr\linewidth-\algorithmicindent}{ Calculating the distillation loss function based on (12).}
        \State \parbox[t]{\dimexpr\linewidth-\algorithmicindent}{Update $\alpha$ by $\mathcal L$.}
    \Until{$t = T$}
    \end{algorithmic}
\end{algorithm}

\section{Property Analysis}
In this section, we analyze the properties of DynaSC in terms of end-to-end transmission time, energy consumption, and RL control system reward. 

\subsection{End-to-end Transmission Time}
Compared with traditional communication, the proposed DynaSC transmits fewer bits for the same source information, thus reducing the transmission delay. However, due to the additional computing time induced by semantic encoding and decoding, the superiority of overall end-to-end transmission time needs to be carefully examined. 
The end-to-end transmission time in DynaSC consists of computing time and transmission time. Inspired by \cite{cai2017neuralpower}, we calculate the computing time of each layer in the neural network separately. For the layer $l$, the computing time ${t_l}(\mathbf {w}_t)$ can be expressed in a polynomial format as
\begin{align}
{t_l}(\mathbf {w}_t)=&\sum_{j}c_{j}\cdot\sum_{k=1}^{n}f(\delta_t)\prod_{i=1}^{D_{l}}a_{ik}^{q_{ijk}}+\sum_{s}c_{s}^{\prime}\sum_{k=1}^{n}f(\delta_t)\mathcal{F}_{s}(\mathbf{a}_{k}), \nonumber\\
&\mathbf{a}_{k}\in\mathbb{R}^{D_{l}}, ~q_{ijk}\in\mathbb{N}, ~\sum_{i=1}^{D_{l}}q_{ijk}\leq K_{l}, ~ \forall j,
\end{align}
where $\mathbf{a}_{k}$ is the $k$-th input vector of the layer $l$ and $a_{ik}$ is its $i$-th element. $q_{ijk}$ is the exponent for $a_{ik}$ in the $j$-th polynomial term, and $c_j$ is the coefficient. $\mathcal{F}_{s}$ is the function related to the hardware specifications associated with layer $l$, including the total number of memory accesses and the total number of floating-point operations, and $c_s$ is the coefficient of the $s$-th polynomial term. In particular, we denote $f(\delta_t)$ as the neural network prediction module. For a subgraph $w_{tk}$ of source $\mathbf w_t$, the output of its prediction module network is denoted as a random variable $\hat \zeta_{tk}$. Then $f(\delta_t)$ can be denoted as
\begin{equation}
\begin{aligned}
f(\delta_t)=\begin{cases}1,\quad\text{if } {\hat\zeta_{tk}}\le\delta_t,\\0,\quad\text{if }{\hat \zeta_{tk}}>\delta_t.\cr\end{cases}
\end{aligned}
\end{equation}
The distribution of $\hat \zeta_{tk}$ can be obtained during neural network training according to the specific control task, and we denote the probability density function of $\hat \zeta_{tk}$ as $p(\hat \zeta_{tk})$. 

For a network with $\mathcal L$ layers, the total computing time is denoted by
\begin{equation}
\begin{aligned}
{T}_{c}(\mathbf {w}_t)=\sum_{l=1}^\mathcal L{{t}_l(\mathbf {w}_t)}.
\end{aligned}
\end{equation}
Given the source information $\mathbf w_t$, the expectation of the computing time is calculated as
\begin{align}
\mathbb E[{T_c}(\mathbf {w}_t)]
=&\sum_{l=1}^\mathcal L\mathbb E{[f(\delta_t)]}\sum_{j}c_{j}\cdot\sum_{k=1}^{n}\prod_{i=1}^{D_{l}}a_{ik}^{q_{ijk}} \nonumber\\
&+\sum_{s}c_{s}^{\prime}\sum_{k=1}^{n}\mathcal{F}_{s}(\mathbf{a}_{k}),
\end{align}
where 
\begin{equation}
\begin{aligned}
\mathbb E[f(\delta_t)]=P(X<k)=\int_{0}^{\delta_t}p(\hat \zeta_{tk})\mathrm{d}\hat \zeta_{tk}.
\end{aligned}
\end{equation}

Next, the transmission time for sending $\mathbf s_t$ is
\begin{equation} 
T_{d}=\frac{Z(\mathbf s_t)}{C},
\end{equation}
where $Z(\mathbf s_t)$ is the data size of $\mathbf s_t$, and $C$ is the channel capacity. Thus, the expectation of the total time for transmitting $\mathbf {w}_t$ can be denoted as
\begin{equation}
\begin{aligned}
\mathbb E[{T}(\mathbf {w}_t)]=\mathbb E[{T_c}(\mathbf {w}_t)]+\frac{Z(\mathbf s_t)}{C}.
\end{aligned}
\end{equation}
The expectation of time reduced by DynaSC, compared with traditional communication, is calculated as
\begin{equation}
\begin{aligned}
\mathbb E[\Delta {T}(\mathbf {w}_t)]=\frac{Z(\mathbf w_t)}{C}-\mathbb E[{T}(\mathbf {w}_t)].
\end{aligned}
\end{equation}

\begin{proposition}
Consider $\gamma$ as the random variable representing the Signal-to-Interference-plus-Noise Ratio (SINR) under a lunar lander to satellite communication link. Due to low elevation angles and scattering caused by lunar surface dust, the channel experiences Rician fading. Under such conditions, the probability that DynaSC reduces the end-to-end transmission time is given by
\begin{equation}
P(\gamma < e^{G_t} - 1) = \int_0^{e^{G_t} - 1} p_\gamma(\gamma)\, d\gamma,
\end{equation}
where the probability density function (PDF) of $\gamma$ is
\begin{equation}
\begin{aligned}
p_\gamma(\gamma) &= (K+1) \cdot \left( \frac{N_0}{\bar{E}_b} \right) \cdot e^{- (K+1) \frac{\gamma N_0}{\bar{E}_b} - K } \\
&\quad \cdot I_0 \left( 2 \sqrt{ K(K+1) \cdot \frac{\gamma N_0}{\bar{E}_b} } \right),
\end{aligned}
\end{equation}
and $I_0(\cdot)$ is the zeroth-order modified Bessel function of the first kind.
\end{proposition}
\renewcommand\qedsymbol{$\blacksquare$}

\begin{proof}
To analyse the probability of DynaSC reducing the end-to-end transmission time, we first introduce a channel model on the lunar surface.
In the lunar environment, the low elevation angle between the lander and the satellite leads to significant multipath effects due to surface reflections. The received signal includes a dominant line-of-sight (LOS) component and a diffuse multipath component, resulting in Rician fading. We refer to the modelling of such lunar surface channels in \cite{divsalar2022comparing} and model the complex baseband channel gain as
\begin{equation}
F(t) = (A \cos(2\pi f_d t) + N_I(t)) + j (A \sin(2\pi f_d t) + N_Q(t)),
\end{equation}
where $A$ is the amplitude of the LOS path, $f_d$ is the Doppler shift, and $N_I(t), N_Q(t)$ are zero-mean Gaussian random processes with equal variance $\sigma_F^2$, representing the diffuse component.

The magnitude of the fading process is $\rho(t) = |F(t)|$, which follows a Rician distribution. Its squared envelope $\rho^2(t)$ scales the received signal power, and the instantaneous SINR can be expressed as:
\begin{equation}
\gamma = \rho^2 \cdot \frac{\bar{E}_b}{N_0},
\end{equation}
where $\bar{E}_b$ is the average bit energy and $N_0$ is the noise power spectral density.

Let us now consider the condition for DynaSC to reduce the end-to-end transmission time. To ensure a reduction in the end-to-end transmission
time, we have
\begin{equation}
\mathbb{E}[\Delta T(\mathbf{w}_t)] > 0.
\end{equation}
Substituting (20) into (21) yields:
\begin{equation}
\frac{Z(\mathbf{w}_t) - Z(\mathbf{s}_t)}{\mathbb{E}[T_c(\mathbf{w}_t)]} > B \log_2(1 + \gamma).
\end{equation}
Let
\begin{equation}
G_t = \frac{\ln 2 \cdot (Z(\mathbf{w}_t) - Z(\mathbf{s}_t))}{B \cdot \mathbb{E}[T_c(\mathbf{w}_t)]},
\end{equation}
Substituting (27) into (28) and using (26), then $\gamma$ must satisfy:
\begin{equation}
\gamma < e^{G_t} - 1.
\end{equation}

As ${\mathbf w}_t$, ${\mathbf s}_t$, and $\delta_t$ are predefined, $G_t$ is only related to $\hat\delta_{tk}$. Since $\gamma$ is defined as a scaled squared Rician variable, we derive its PDF using change of variables. The PDF of $\rho$ is
\begin{equation}
p(\rho) = 2(K+1) \rho e^{-(K+1)\rho^2 - K} I_0\left(2\rho \sqrt{K(K+1)}\right),
\end{equation}
and using $\gamma = \rho^2 \cdot \bar{E}_b / N_0$, the PDF of $\gamma$ becomes:
\begin{equation}
\begin{aligned}
p_\gamma(\gamma) &= (K+1) \cdot \left( \frac{N_0}{\bar{E}_b} \right) \cdot e^{- (K+1) \frac{\gamma N_0}{\bar{E}_b} - K } \\
&\quad \cdot I_0 \left( 2 \sqrt{ K(K+1) \cdot \frac{\gamma N_0}{\bar{E}_b} } \right),
\end{aligned}
\end{equation}

Finally, the probability that the SINR lies below the threshold and hence DynaSC achieves transmission time improvement is:
\begin{equation}
P(\gamma < e^{G_t} - 1) = \int_0^{e^{G_t} - 1} p_\gamma(\gamma) \, d\gamma.
\end{equation}
\end{proof}

\subsection{Energy Consumption}
The considered lunar landing control scenario is energy-constrained and the communication modules have an independent energy supply. Therefore, it is meaningful to measure energy consumption in the DynaSC scheme, which includes the energy consumption of coding and data transmission. We use a similar way to calculate the computing energy consumption of the neural network as follows:
\begin{align}
E_l(\mathbf {w}_t)&\!=\!\! \sum_jd_j\cdot\sum_{k=1}^{n}f(\delta_t)\prod_{i=1}^{D_l}a_{ik}^{m_{ijk}} \!+\! \sum_rd_r'\sum_{k=1}^{n}f(\delta_t)\mathcal{F}_r(\mathbf{a}_k) \nonumber\\
&\mathbf{a}_k\in\mathbb{R}^{D_l}, ~ m_{ijk}\in\mathbb{N}, ~ \sum_{i=1}^{D_l}m_{ijk}\leq K_l', ~\forall j,
\end{align}
where $d_j$ is the coefficient of the $j$-th polynomial term and $d'_r$ is the coefficient of the $s$-th polynomial term. For a network with $\mathcal{L}$ layers, the total computing energy consumption is
\begin{equation}
\begin{aligned}
{E}_c(\mathbf {w}_t)=\sum_{l=1}^\mathcal{L}{{E}_l(\mathbf {w}_t)}.
\end{aligned}
\end{equation}
The transmission energy for sending $\mathbf s_t$ is
\begin{equation} 
E_{d}=y_{d}(\mathbf s_t).
\end{equation}
Thus, the total communication and computation energy consumption for transmitting $\mathbf {w}_t$ of the system is
\begin{equation}
\begin{aligned}
E=\sum_{l=1}^\mathcal L{{E}_l}+{{y}_{d}}({{\mathbf{s}}_{t}}).
\end{aligned}
\end{equation}

\subsection{The Belief Property of DynaSC}
SemCom can accurately recover received information through neural network-based semantic coding, particularly under poor channel conditions where syntax errors are severe. However, when the channel quality is high and syntax errors are minimal, the advantages of SemCom over traditional communication schemes become less apparent. This is due to potential semantic ambiguities introduced by the neural network and background knowledge, which may degrade the accuracy of recovered information. Consequently, we analyze the channel conditions under which the proposed DynaSC achieves performance gains in transmission accuracy.

For a given $y_t$, the observations of traditional communication and DynaSC are denoted as $z_{t,g}$ and $z_{t,s}$, respectively and computed as follows:
\begin{align}
z_{t,g}&=\eta\left( h(t)*w_t+n \right), \\
z_{t,s}&=\eta\left( \mathcal{ E}_d (h(t)*\mathcal{ E}_e(w_t; { \boldsymbol \alpha})+n; { \boldsymbol \alpha}) \right).
\end{align}
Given SNR $\gamma= \mathbb E[h(t)^2]/ \mathbb E[n^2]$, under the observations of $z_{t,g}$ and $z_{t,s}$, the corresponding beliefs $b_{t,g}$ and $b_{t,s}$ can be represented as the distribution $p(b_{t,g}|x_t,y_t,\gamma)$ and $p(b_{t,s}|x_t,y_t,\gamma)$, respectively. According to (3) and (4), the actions $u_{t,g}$ and $u_{t,s}$ are taken based on $b_{t,g}$ and $b_{t,s}$, respectively. According to the distribution of $b_{t,g}$ and $b_{t,s}$, we denote the probability density function of $u_{t,g}$ and $u_{t,s}$ as $p(u_{t,g}|p(b_{t,g}|x_t,y_t,\gamma))$ and $p(u_{t,s}|p(b_{t,s}|x_t,y_t,\gamma))$, respectively.
Hence,  the corresponding expectation of reward is denoted by
\begin{align}
\mathbb E[R_g]&=\sum_{u_{t,g}\in \mathcal A}p(u_{t,g}|p(b_{t,g}|x_t,y_t,\gamma)) \cdot r(b_{t,g},u_{t,g}), \\
\mathbb E[R_s] &=\sum_{u_{t,s}\in \mathcal A}p(u_{t,s}|p(b_{t,s}|x_t,y_t,\gamma)) \cdot r(b_{t,s},u_{t,s}).
\end{align}

We evaluate the effectiveness of the communication policy for control based on the expected reward. When $\gamma$ makes $\mathbb E[R_s]>\mathbb E[R_g]$, SemCom can achieve a better effectiveness in the considered RL control system.


\begin{table*}[t]
    \setlength{\abovecaptionskip}{-5pt}
    \setlength{\belowcaptionskip}{-25pt}
    \vspace{-1mm}
    \small
    \caption{High-resolution Lunar Surface Images Dataset}
    \begin{center}
        \renewcommand{\arraystretch}{1.2}
        \setlength{\tabcolsep}{8pt}
        \begin{tabular}{c c c c}
            \toprule[1.5pt]
            \textbf{Dataset Name} & \textbf{Type of Images} & \textbf{Dimensions}& \textbf{Number of segmented images} \\
            \midrule[1pt]
            M1447547309R   & Luna 25 Impact Crater           &     $5765 \times 14073$          & 6084  \\
            \addlinespace[0.5em]
            M1447750764LR  & Chandrayaan-3 Landing Site            &      $20000 \times 20000$    & 6422  \\
            \addlinespace[0.5em]
            M1173350480L/R & Gruithuisen Collapsed Lava Tube        &       $3700 \times 6100$   & 1188  \\
            \addlinespace[0.5em]
            M1472410644L   & Chang'e 6 sample return spacecraft first look  & $43361 \times 9864$ & 11264 \\
            \bottomrule[1.5pt]
        \end{tabular}
        \label{tab2}
    \end{center}
\end{table*}

\begin{figure*}[htp]
	\centering
        \setlength{\abovecaptionskip}{0pt}
        \setlength{\belowcaptionskip}{0pt}
	\includegraphics[width=0.9\textwidth]{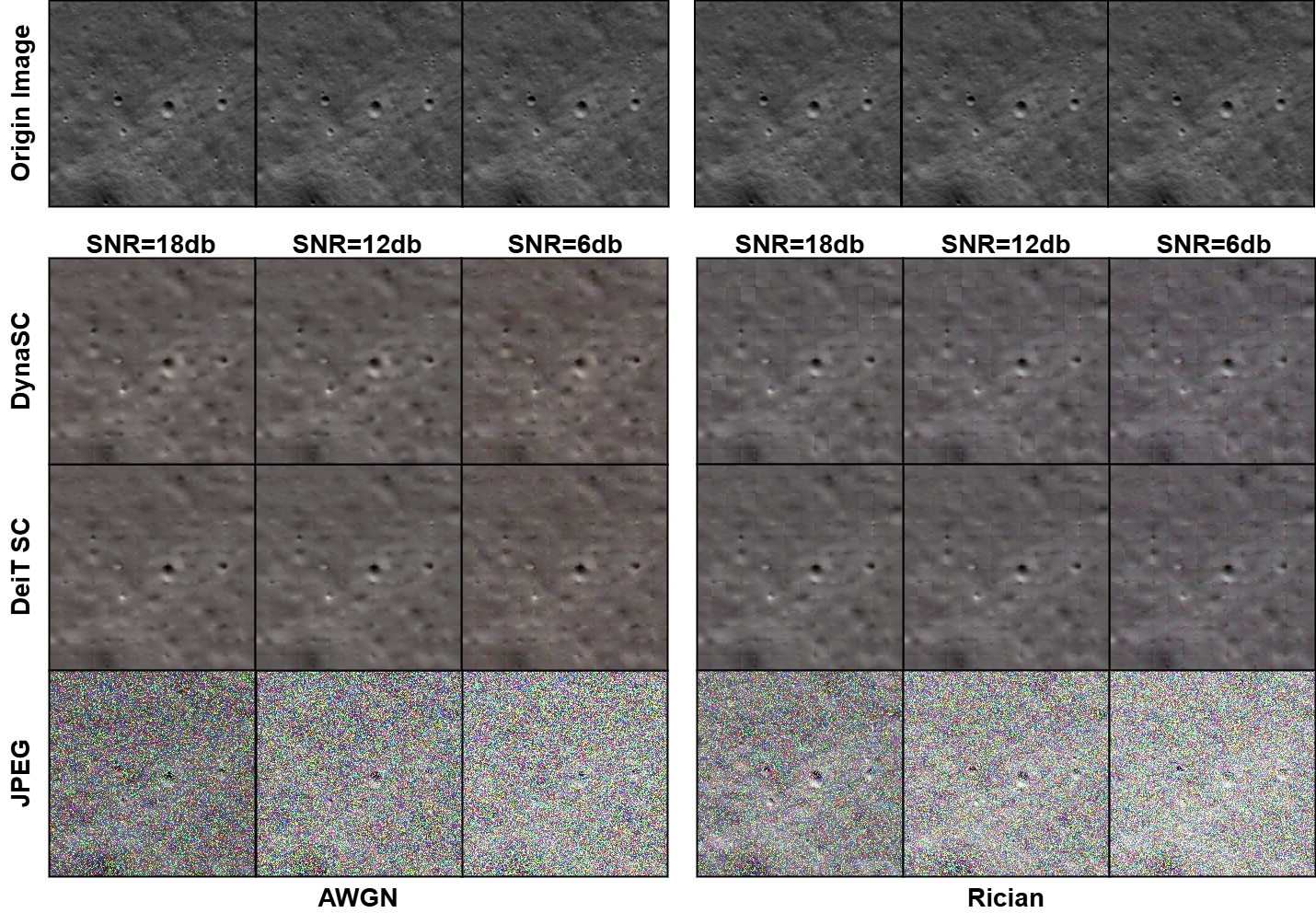}
	\caption{Comparison of DynaSC, DeiT-SC and JPEG reconstructed lunar surface images with the original images under different SNR.(Flat lunar surface)}
	\label{img2}
\end{figure*}
\section{Simulation Results and Discussions}
We conduct simulations at two distinct levels. First, we evaluate the transceiver performance of the proposed DynaSC in terms of accuracy and latency. Next, we assess the landing performance associated with the landing control-aware SemCom framework.

\subsection{Simulation Setup}
\subsubsection{Landing Control-aware SemCom System Setup}
For training the proposed DynaSC transceiver, we utilize four ultra-high-resolution lunar surface images obtained from the Lunar Reconnaissance Orbiter (LROC) database (http://wms.lroc.asu.edu/lroc): M1447547309R, M1447750764LR, M1173350480L/R, and M1472410644L. Each image captures a large-scale overhead view of the lunar surface, with spatial coverage ranging from a few kilometers to several tens of kilometers. These images include detailed representations of diverse lunar terrains such as impact craters, collapsed lava tubes, and smooth plains—features that are directly relevant to key tasks in lunar landing, including site detection and hazard avoidance.
To enable training at a fine semantic level, each image is segmented into non-overlapping patches of size 256 × 256 pixels, yielding a total of 24,958 training samples. These patches collectively represent a wide range of illumination conditions, terrain geometries, and scale variations, allowing the model to learn from realistic and richly structured surface data. Among all patches, 95\% are used for training and the remaining 5\% for testing. Detailed statistics including scene type, number of patches, and spatial resolution are provided in Table II.
\begin{figure*}[htp]
	\centering
        \setlength{\abovecaptionskip}{0pt}
        \setlength{\belowcaptionskip}{0pt}
	\includegraphics[width=0.9\textwidth]{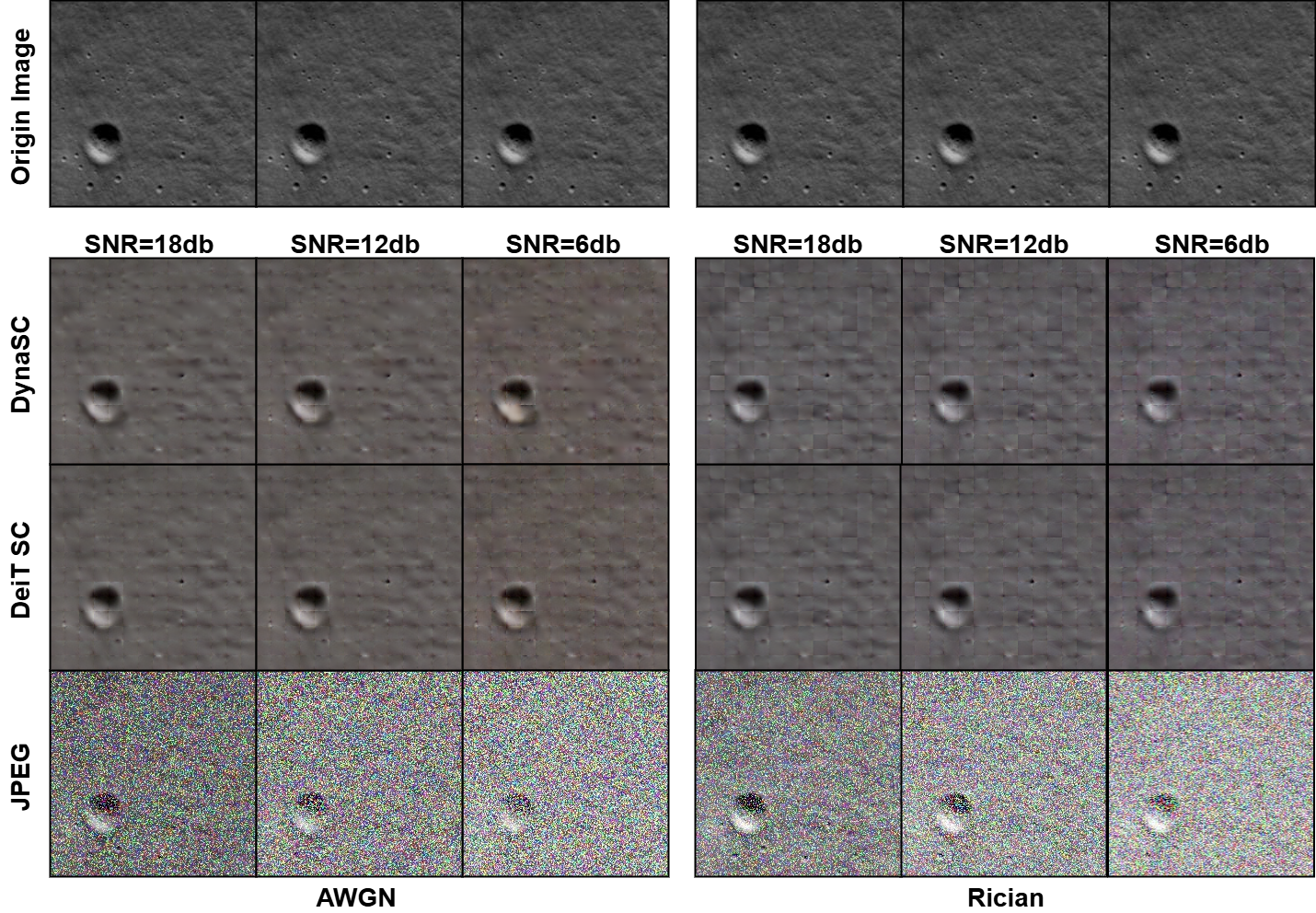}
	\caption{Comparison of DynaSC, DeiT-SC and JPEG reconstructed lunar surface images with the original images under different SNR.(Crater of impact)}
	\label{img2}
\end{figure*}
\begin{table}[t]
    \setlength{\abovecaptionskip}{2pt}
    \setlength{\belowcaptionskip}{0pt}
    \small
    \renewcommand{\arraystretch}{1.2}
    \centering
    \caption{Key simulation parameters.}
    \label{tab:sim_params}
    \begin{tabular}{|p{3.0cm}|p{4.8cm}|}
        \hline
        \textbf{Parameter} & \textbf{Value / Description} \\
        \hline
        Encoder transformer & 12 layers, 12 heads \\
        \hline
        Decoder transformer & 8 layers, 12 heads \\
        \hline
        Training $\delta_t$ & 0.7 (fixed) \\ 
        \hline
        Batch size & 16 \\
        \hline
        Channel model & Rician fading, $K=5$ \\
        \hline
        RL environment & $6\,\text{km} \times 2\,\text{km} \times 2\,\text{km}$ \\
        \hline
        RL episodes & 20,000 \\
        \hline
    \end{tabular}
\end{table}

In the model training, we adopt a transformer-based encoder–decoder architecture, and the training is carried out using Adam optimization\cite{adam} with the learning rate ${10}^{-4}$, and the batch size is 16. All the codes are implemented in Pytorch and conducted on an RTX A6000 GPU. All the key simulation and training parameters are summarized in Table~\ref{tab:sim_params}.

We compare our method with two benchmarks: DeiT SemCom (DeiT-SC) model\cite{DeiT} (without dynamic sparsity module), and the common engineering code JPEG. For a fair comparison, all the structures of the DeiT-SC, except the prediction module, are the same as those of DynaSC with the same parameter settings. 

We evaluate the performance of landing site detection during a lunar landing mission for the three transceivers. Specifically, we calculate the Peak Signal-to-Noise Ratio (PSNR) between each segmented patch and the reconstructed image patch, identifying the patch with the highest PSNR value. We then determine the probability that this highest-PSNR patch matches the original pre-transmission image patch. Through this process, we assess the impact of DynaSC on the lunar lander’s navigation performance.

Given the long distance between the lunar orbiting satellite and the lunar lander, the lander’s relative position to the satellite can be assumed to be fixed. Consequently, we consider the SNR to remain constant throughout the landing process. However, to account for varying landing scenarios, the SNR is treated as an adjustable parameter under different conditions.
To reflect the non-ideal nature of lunar surface communications, our simulations adopt a Rician fading channel model. This setting aligns with the lunar surface communication model discussed in \cite{divsalar2022comparing}, where low elevation angles and reflective surface effects result in Rician fading with both dominant and scattered components. To simulate different levels of environmental degradation, we vary the SNR to account for thermal noise variation, surface scattering from lunar dust, and Doppler shifts due to lander motion. These channel conditions allow us to evaluate the robustness of DynaSC under realistic and degraded transmission scenarios.

\subsubsection{Lunar Landing System Setup}
To conduct numerical simulations of a lunar surface landing system, we define a space measuring 6 km in length, 2 km in width, and 2 km in height. Within this space, the lander is equipped with a single-hole camera, oriented perpendicular to the lunar surface, to capture images of the surface. Additionally, it has six propulsion modules that enable motion along all three axes of the lander’s body. The lunar surface is divided evenly into a set of 30×10 regions.
Prior to the initial phase of the landing mission, a target landing zone is assigned randomly. The lander then enters a free-fall state with a predetermined initial velocity and begins its maneuvering process after a specified number of steps. The lander’s actions are controlled by a trained RL agent. The agent receives noisy images from the wireless transmission, which provide observations of the target landing zone direction. Based on these observations, the agent selects an action and computes a corresponding reward to update the policy.

In particular, we organize the post-touchdown states of the lander on the lunar surface into three categories: perfect landing, imperfect landing, and landing failure. A perfect landing indicates that the lander reaches the designated landing site precisely. An imperfect landing is that the lander does not reach the target site but remains intact. In contrast, a landing failure indicates that the lander crashes.


\setcounter{figure}{4}
\begin{figure}[t]
	\centering
         \setlength{\abovecaptionskip}{0pt}
        \setlength{\belowcaptionskip}{0pt}
	\includegraphics[width=0.9\linewidth]{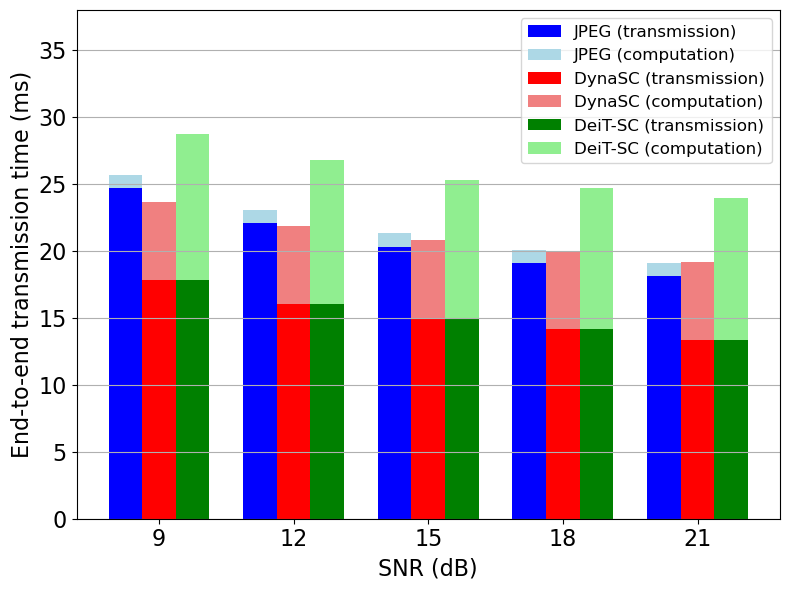}
    \vspace{-2mm}
	\caption{The average transmission and computing time under different SNRs.}
	\label{img3}
\end{figure}
\begin{figure}[t]
	\centering
        \setlength{\abovecaptionskip}{0pt}
        \setlength{\belowcaptionskip}{0pt}
	\includegraphics[width=0.45\textwidth]{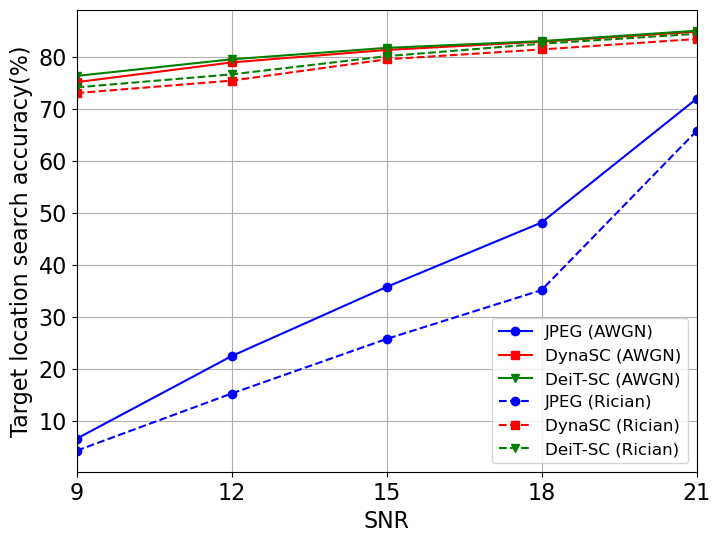}
    \vspace{-2mm}
	\caption{The Landing site search accuracy under different SNR.}
	\label{img2}
\end{figure}
\setcounter{figure}{6}
\begin{figure}[t]
	\centering
        \setlength{\abovecaptionskip}{-0pt}
        \setlength{\belowcaptionskip}{-0pt}
	\includegraphics[width=0.45\textwidth]{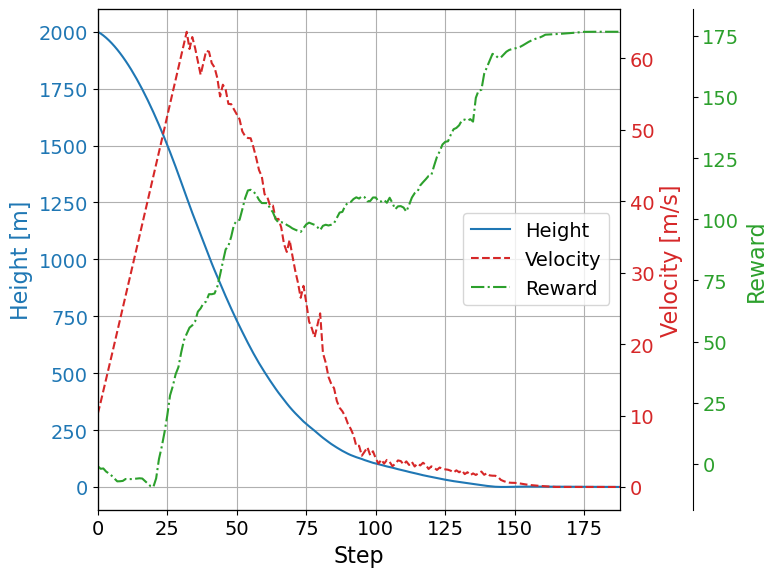}
    \vspace{1mm}
	\caption{The velocity, height, and reward vs. landing steps based on DynaSC.}
	\label{img3}
\end{figure}
\setcounter{figure}{7}
\begin{figure*}[t]
	\centering
        \setlength{\abovecaptionskip}{0pt}
        \setlength{\belowcaptionskip}{0pt}
	\includegraphics[width=1\textwidth]{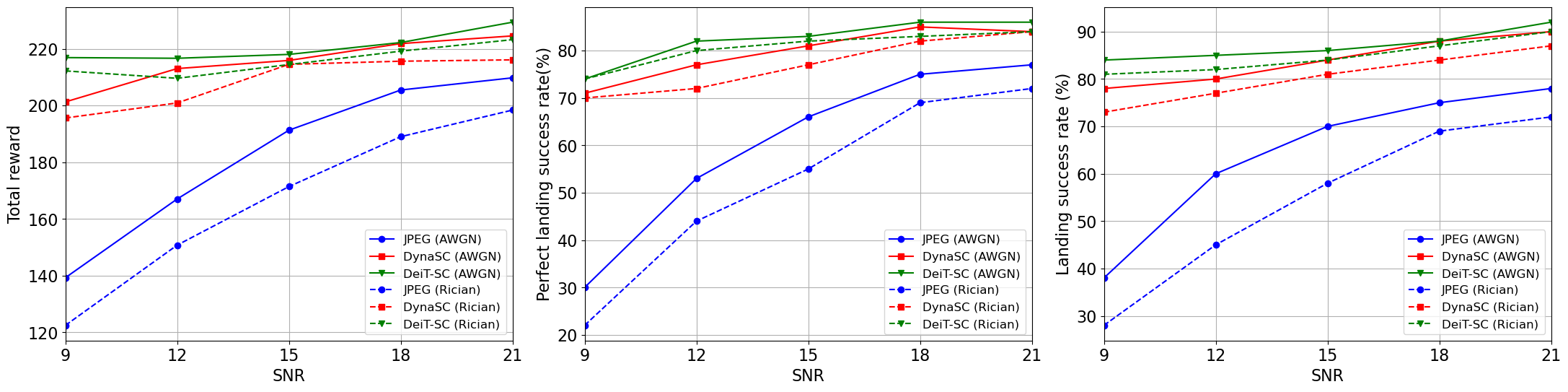}
	\caption{The numerical lunar landing simulation performance of the DynaSC and two benchmarks.}
	\label{img4}
\end{figure*}

\subsection{DynaSC Transceiver Performance}
In the following, we evaluate DynaSC's performance against two benchmarks in terms of lunar surface remote sensing image recovery, landing site search accuracy, and end-to-end transmission time.

Fig. 3 and Fig. 4 presents a comparison between the reconstructed lunar surface images and the original images using the three transmission methods under various SNR levels in AWGN and Rice fading channels. 
The AWGN channel serves as a thermal noise baseline, while the Rician fading channel introduces both line-of-sight and multipath components caused by lunar surface reflections and scattering, offering a more realistic model of lunar communication impairments. We select a featureless image of the lunar surface and another featuring a prominent crater as sample images for demonstration. 
Under AWGN, it is evident that both DynaSC and DieT-SC outperform JPEG in preserving the original image features, successfully reconstructing major craters and landform details even at SNR = 6. 
When tested under the more challenging Rician fading conditions, JPEG’s performance drops sharply, losing almost all semantic detail. In contrast, the two SemCom-based methods remain robust, preserving key structural features across SNR levels. 
This is because the SemCom transceiver mitigates the impact of noise on the overall image semantics by reconstructing lost information using data from adjacent regions surrounding the noisy area. 
Notably, DynaSC achieves reconstruction quality comparable to DeiT-SC while reducing both transmission and computation overhead. By dynamically prioritizing semantically important patches, DynaSC maintains the performance of high-capacity models without incurring their full computational cost, confirming its effectiveness under degraded lunar channel conditions.

Next, we evaluate the end-to-end transmission time of the three transceivers under varying SNR conditions, using a bandwidth of 3 MHz. Each result is averaged over 100 trials, with each trial transmitting a 32 Kb lunar surface image. The results are shown in Fig. 5, where the total time is decomposed into transmission time and computation time via stacked bar charts.
As expected, the overall end-to-end time for all methods decreases as SNR increases, owing to the higher channel capacity. Notably, DynaSC consistently achieves the shortest total latency under low SNR conditions, with a clear advantage in both components: it transmits fewer semantic bits and reduces computational load by selectively encoding only semantically critical patches. This dual efficiency is reflected in the noticeably smaller computational time segment compared to DeiT-SC.
In contrast, JPEG exhibits negligible computation time but transmits significantly more data, resulting in longer transmission time under low SNR. However, as the SNR increases, the end-to-end transmission time of JPEG becomes nearly identical to that of DynaSC, which coincides with our property analysis of DynaSC presented in Section V.

We then examine the impact of DynaSC on the lunar lander's navigation performance. As shown in Fig. 6, the accuracy of landing site detection increases with SNR for all three transceiver methods under both AWGN and Rician fading channels.
Under the AWGN channel, both DynaSC and DeiT-SC achieve higher detection accuracy than JPEG across all SNR levels, and the accuracy of the two SemCom-based methods is nearly the same.
In the Rician fading channel, the detection accuracy becomes lower for all methods, especially when the SNR is low. JPEG drops significantly in performance, while DynaSC still keeps a high level of accuracy that is close to DeiT-SC. This shows that DynaSC remains reliable even under more difficult channel conditions.
This is because, although DynaSC reduces the encoding computation for non-critical regions, the features in these areas are not highly distinct. As a result, even without high-performance neural network encoding, these areas can still be reconstructed using information from adjacent regions. For critical image regions, DynaSC applies the same high-quality encoding as DeiT-SC, resulting in nearly identical reconstruction outcomes.
These results confirm that DynaSC preserves the original model's performance while reducing overall computational cost, and remains effective even under degraded channel conditions representative of lunar environments.

\subsection{Lunar Landing Performance}
We evaluate the performance of DynaSC in the lunar landing scenario. Fig. 7 shows the relationship between the velocity, height, and reward of the lunar lander with respect to landing steps using DynaSC.
We observe that the velocity increases rapidly during the free-fall phase, then decreases as the lander begins its maneuvers, gradually approaching zero as it nears the ground. The height decreases with each step, ultimately reaching zero, while the reward increases with each step and stabilizes after a successful landing. These results align with expectations, confirming the effectiveness of our proposed DynaSC-based landing framework.

In Fig. 8, we compare the landing performance of DynaSC with two benchmarks. From Figs. 8(a), 8(b), and 8(c), we can observe that the total reward, perfect landing success rate, and landing success rate of DynaSC and the benchmarks all increase with SNR. 
For all three methods, performance improves with increasing SNR in both channel conditions. Under the AWGN channel, DynaSC and DeiT-SC achieve higher values across all metrics compared to JPEG, with minimal difference between them. When switching to the more challenging Rician fading channel, the overall performance of all methods drops—most notably for JPEG, which suffers sharp declines at low SNR. In contrast, both DynaSC and DeiT-SC show relatively stable performance across channels, highlighting the robustness of SemCom in degraded lunar communication environments.
Moreover, the gap between the SemCom methods and JPEG becomes larger under the Rician channel, especially at low SNR. Despite the presence of stronger noise and multipath fading, DynaSC maintains landing success rates and total rewards close to those of DeiT-SC.
This shows that SemCom methods can still support reliable control decisions in degraded channel conditions by transmitting the key visual information needed for landing.

\section{Conclusion}
In this paper, we explored a SemCom-based lunar landing control framework, where a novel SemCom transceiver, DynaSC, is developed. By dynamically adjusting the transmission strategy, DynaSC ensures the reliable delivery of critical information required for lunar landing across various channel conditions. 

This work provides pioneering insights into the research of SemCom in space communications. The information recovery capability of SemCom will be instrumental in offering critical support under the extreme channel conditions and resource constraints present in specific space communication scenarios.
Moreover, this work offers valuable insights for the investigation of the joint design of SemCom and control systems, such as dynamically adjusting the encoding strategy of SemCom based on the critical characteristics of the control system. Looking forward, future research may further explore the integration of multi-modal sensing data, the joint design of SemCom with satellite edge computing platforms, and the extension of DynaSC to multi-agent coordination scenarios. These directions will enhance the robustness, scalability, and practicality of SemCom frameworks for deep-space missions.


%

\appendices
\section{Proof of Proposition 1}
Let $z_{t1}$ and $z_{t2}$ be the observations at times $t_1$ and $t_2$, with the corresponding observed distributions are $\Omega(z_{t_1})$ and $\Omega(z_{t_2})$, respectively. 
The variances of $\Omega(z_{t_1})$ and $\Omega(z_{t_2})$ are denoted by $\sigma_{t_1}$ and $\sigma_{t_2}$, respectively.
When the agent is more certain about the environment, $\Omega(z_{t_1})$ and $\Omega(z_{t_2})$ will be concentrated near the true environment, as shown in Fig. 9. Therefore, when $\frac{{{\partial }}{r(b_{t_1},u_{t_1})}}{\partial u_{t_1}}>\frac{{{\partial }}{r(b_{t_2},u_{t_2})}}{\partial u_{t_2}}$, we have $\sigma_{t_1}>\sigma_{t_2}$.
\setcounter{figure}{8}
\begin{figure}[htp]
	\centering
        \setlength{\abovecaptionskip}{0pt}
        \setlength{\belowcaptionskip}{0pt}
	\includegraphics[width=0.3\textwidth]{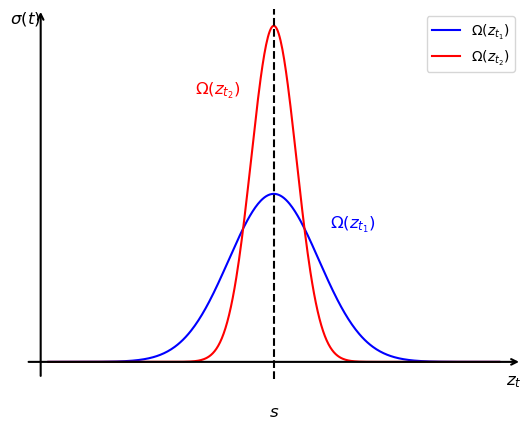}
	\caption{The observed distributions $\Omega(z_{t_1})$ and $\Omega(z_{t_2})$.}
	\label{img3}
\end{figure}

Let $\delta=\delta-\theta$ at time $t_1$ and $t_2$. In order to find the moment when reducing $\delta$ has less loss of accuracy on RL observations, we assume that the variances of $\Omega(z_{t_1})$ and $\Omega(z_{t_2})$ are $\sigma_{t_1}+{\epsilon}^2$ and $\sigma_{t_2}+{\epsilon}^2$, respectively. Given that $\sigma_{t_1}>\sigma_{t_2}$, the rate of change of the variances satisfy $\frac{{\epsilon}^2}{\sigma_{t_1}+{\epsilon}^2}<\frac{{\epsilon}^2}{\sigma_{t_2}+{\epsilon}^2}$.



\ifCLASSOPTIONcaptionsoff
  \newpage
\fi

\bibliographystyle{IEEEtran}
\bibliography{IEEEexample}
\vfill
\begin{IEEEbiography}
[{\includegraphics[width=1in,height=1.25in,clip,keepaspectratio]{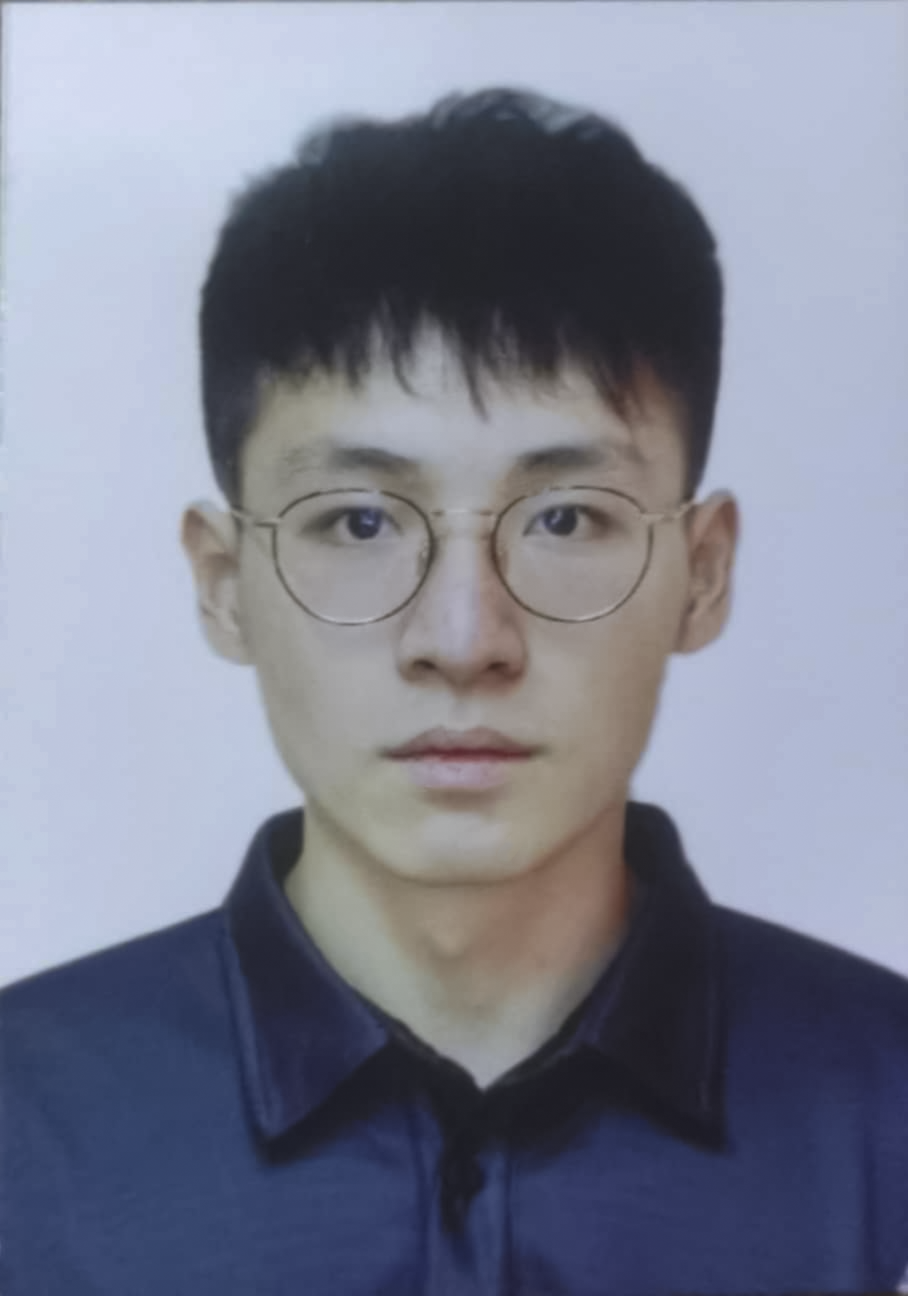}}]
{Fangzhou Zhao} (Graduate Student Member, IEEE) received the B.Eng. degree from the School of Information Engineering, Zhejiang University of Technology, Hangzhou, China, in 2018, and the M.Sc. degree from the James Watt School of Engineering, University of
Glassgow, Glassgow, U.K.,in 2020. He is currently working toward the Ph.D. degree with James Watt School of Engineering, University of Glasgow, Glasgow, U.K. His research interests include deep learning in wireless communication and semantic communication.
\end{IEEEbiography}

\begin{IEEEbiography}
[{\includegraphics[width=1in,height=1.25in,clip,keepaspectratio]{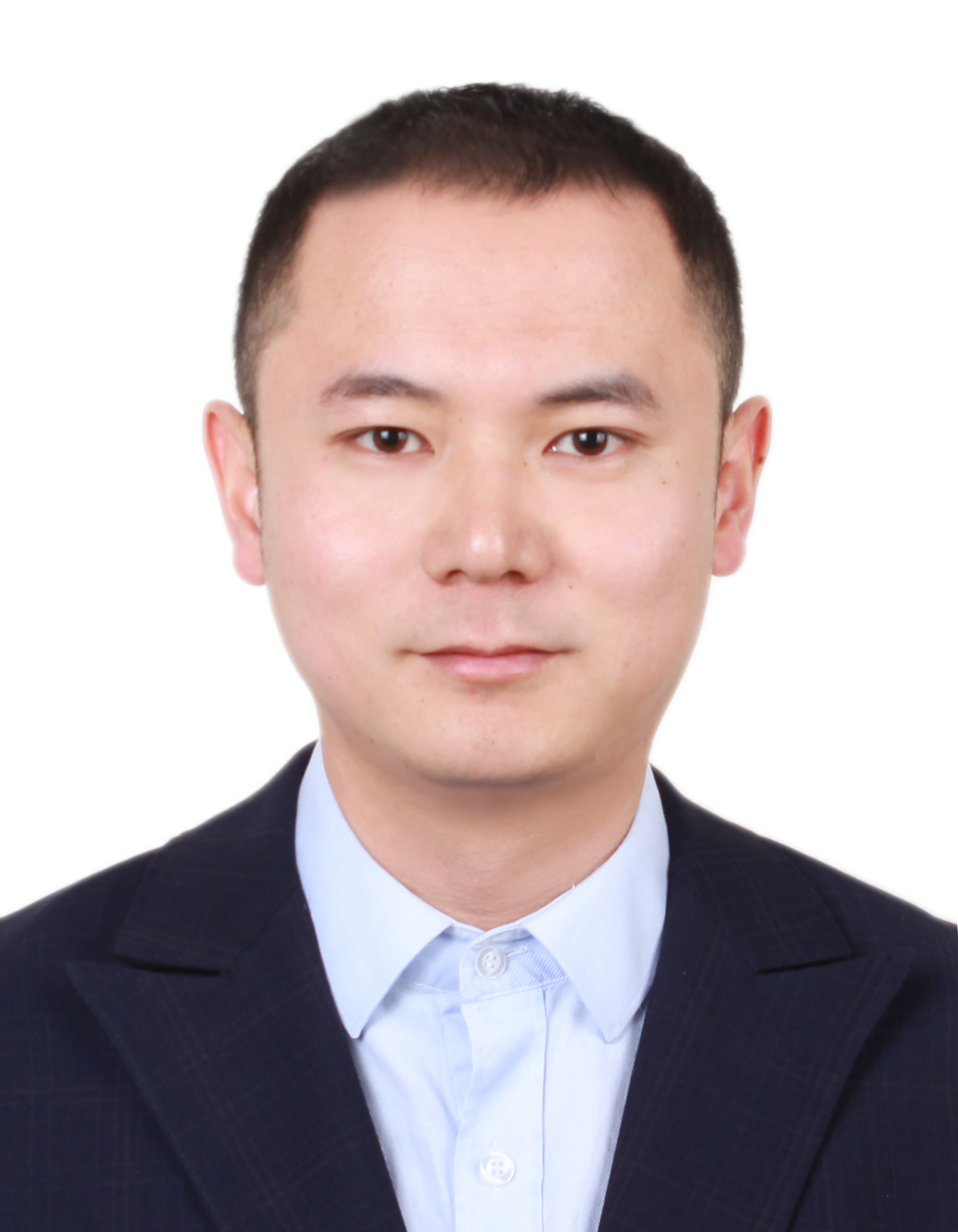}}]
{Yao Sun} (Senior Member, IEEE) is currently a Lecturer with James Watt School of Engineering, the University of Glasgow, Glasgow, UK. Dr Sun has won the IEEE Communication Society of TAOS Best Paper Award in 2019 ICC, IEEE IoT Journal Best Paper Award 2022 and Best Paper Award in 22nd ICCT. His research interests include intelligent wireless networking, semantic communications, blockchain system, and resource management in next generation mobile networks. Dr. Sun is a senior member of IEEE.
\end{IEEEbiography}

\begin{IEEEbiography}
[{\includegraphics[width=1in,height=1.25in,clip,keepaspectratio]{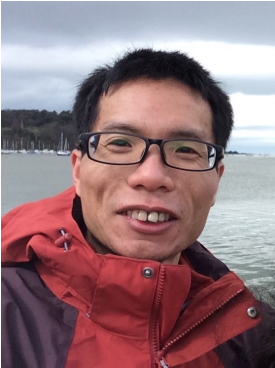}}]
{Jianglin Lan} received the Ph.D. degree from the University of Hull in 2017. He has been a Leverhulme Early Career Fellow and Lecturer at the University of Glasgow since 2022. He was a Visiting Professor at the Robotics Institute, Carnegie Mellon University, in 2023. From 2017 to 2022, he held postdoc positions at Imperial College London, Loughborough University, and the University of Sheffield. His research interests include AI, optimisation, control theory, and autonomy
\end{IEEEbiography}

\begin{IEEEbiography}
[{\includegraphics[width=1in,height=1.25in,clip,keepaspectratio]{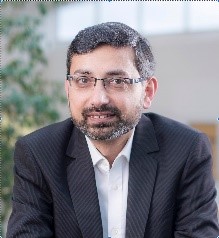}}]
{Muhammad Ali Imran} (Fellow, IEEE) is a Professor of Wireless Communication Systems and Dean of Graduate Studies in College of Science and Engineering. His research interests include self-organized networks, wireless networked control systems, and the wireless sensor systems. He heads the Communications, Sensing and Imaging CSI Hub, University of Glasgow, Glasgow, U.K. He is also an Affiliate Professor with The University of Oklahoma, Norman, OK, USA, and a Visiting Professor with the 5G Innovation Centre, University of Surrey, Guildford, U.K. He has more than 20 years of combined academic and industry experience with several leading roles in multimillion pounds funded projects. His research interests include self-organized networks, wireless networked control systems, and the wireless sensor systems.
\end{IEEEbiography}
\vfill
\end{document}